\documentstyle[floats,multicol,aps,epsf,prb]{revtex}

\begin{document}
\ifx\href\undefined\else\errmessage{hyperTeX disabled by xxx admin}\fi
\draft
\twocolumn[\hsize\textwidth\columnwidth\hsize\csname @twocolumnfalse\endcsname
\title{Thermal currents in highly correlated systems}
\author{J. Moreno and P. Coleman}
\address{Department of Physics and Astronomy, Rutgers University,
Piscataway, NJ 08855}
\date{\today}
\maketitle

\begin{abstract}
Conventional approaches to thermal conductivity in itinerant systems
neglect the contribution to thermal current due to
interactions. We derive this  contribution
to the thermal current and show how it produces
important corrections to the thermal conductivity in 
anisotropic superconductors. We discuss the possible 
relevance of these corrections 
for the interpretation of the thermal conductivity of 
anisotropic  superconductors.
\end{abstract}
\vskip2pc]
\def\vx{\vec x}
\def\vj{\vec j}
\def\vr{\vec r}
\def\vk{\vec k}
\def\vq{\vec q}
\def\vp{\vec p}
\input psfig.tex
Historically, understanding thermal conductivity
has played an important role in the development of
materials theory.
There has recently been a growth
of interest in studying this basic property 
in strongly correlated materials.\cite{Taillefer,recent}   
One of the recurrent consequences of strong interactions 
is the development of anisotropic gaps in the 
excitation spectrum: possible examples  include
heavy fermion superconductors, Kondo insulators and the chevrel
superconductor $V_3Si$.
\cite{hfermion,v3si} In such 
systems, the thermal conductivity has an important role
to play in the elucidation of the gap symmetry.

Pioneering work on the theory of thermal conductivity was carried out
in the sixties\cite{Chester,Langer,super};  by and large, the theoretical 
approaches taken today
are a direct application of this early work \cite{phonon,Tmatrix,Boltzman}.
In this paper, we show how these 
classic approaches require modification 
to take account of new thermal conduction channels
that are introduced by interactions. As part of this
discussion, we shall show how thermal conductivity, like
particle or charge conductivity, can be regarded as a boundary condition
response which is most directly computed as a response to a
fictitious gauge field.  We illustrate this  approach  in the context of 
anisotropic superconductivity, showing how
interaction contributions to the thermal current 
significantly modify the flow of heat produced by excitations
near a gap node.

Thermal conductivity 
is directly related, via the Kubo formula, to 
thermal current fluctuations 
\begin{equation}
\kappa^{ab} = -  \beta \hspace{0.05in} lim_{\omega \rightarrow 0} 
\frac{\partial} {\partial \omega}
Re \{  \int _0^{\infty} dt e^{i \omega t}
\langle [ j^{a}(t), j^{b}(0)]\rangle \}.
\end{equation}
There is an important distinction between heat and charge
conductivity.  Unlike charge density, the energy density does not
commute with the interactions: this means that heat is directly
transmitted by interactions, introducing new interaction contributions
to the microscopic thermal current operator.  The effect of these new
thermal conduction channels is important when interactions severely
modify the dispersion of the electrons, as in the case of a 
Mott insulator, or a superconductor with gap nodes.
\cite{mahan,polaron,castellani} 

To derive the thermal current 
we  appeal to Noether's theorem, which relates
continuity of energy flow to the covariance
of the action under co-ordinate transformations in time
$
t\rightarrow t' = t + \phi[\vec x, t'].
$
Consider an
electronic system described by the Lagrangian
\begin{equation}
{\cal L} = \int d^3x(i\psi^{\dagger}
\!\stackrel{\leftrightarrow}{\partial_t}\!
\psi) -H,
\end{equation}
where $H$ is the Hamiltonian
and 
$\! \stackrel{\leftrightarrow}
{\partial_t} = {1 \over 2}
(\stackrel{\rightarrow}{\partial_t} - 
\stackrel{\leftarrow}{\partial_t})\!$ is the antisymmetrized
time-derivative.
By Noether's theorem, the change in the action 
$S= \int {\cal L}(t) dt$ is the energy continuity equation
\begin{equation}
{\delta S/ \delta \phi[\vec x, t] } = 
-\bigl[\partial _t \epsilon({\vec x},t) + {\vec \nabla}\cdot
{\vec j}({\vec x},t)
\bigr],
\end{equation}
where $\epsilon$  and $\vec j$ are  the energy 
and thermal current density respectively.
Continuity of energy flow follows from the invariance
of the action under co-ordinate transformations 
${\delta S/ \delta \phi[\vec x, t] } = 0$.
Writing $\delta S 
= \int [\delta S/ \delta \phi] \delta \phi(x,t)$ and integrating
by parts, we find
\begin{equation}
\delta S = \int dt d\vec x \biggl[
\dot \phi(\vec x,t) \epsilon(\vec x, t)
+
\vec \nabla \phi(\vec x,t)\cdot \vec j(\vec x, t)
\biggr]
\end{equation}
so that 
$\epsilon(x) = {\delta S / \delta  {\dot  \phi}}$, while the
heat current is ${\vec j} = {\delta S / \delta  {\vec \nabla} \phi}
$.
The 
calculation of the thermal current requires special care
because  interactions  are non-local
and all higher derivatives of the
fields must be taken into account when taking the
functional derivative.  
This important point was overlooked in the classic treatment by Langer
\cite{Langer}.
We can take these effects into account by 
noting that the derivative
operators inside the action are evaluated at constant time $t$, 
so that under the transformation $t\rightarrow t'$, they
acquire a covariant form
\begin{equation}
- i \vec \nabla \rightarrow 
- i {\vec \nabla} - \vec A \hat w,
\end{equation}
where we denote $\hat \omega= i\!\stackrel{\leftrightarrow}{\partial_{t'}}$.
The field
$\vec A = \vec \nabla \phi$ 
adjusts for the fact that spatial derivative $\nabla \equiv\nabla\vert_{t'}$
is
taken at constant $t'$. It follows that 
the energy current is 
\begin{equation}
\vec j = {\delta S
/\delta \vec \nabla \phi } = -{\partial H[ \vec A] / \partial \vec A},
\end{equation}
where $H[\vec A]$ is the ``gauged'' Hamiltonian with the 
replacement $\vec k \rightarrow \vec k - \vec A \hat \omega$ 
in both the kinetic {\em and} interaction terms. 
We see that $\vec A$ plays the role of 
a ``fictitious'' gauge
field conjugate to the thermal current.  

To illustrate  this procedure, consider 
a fluid of electrons with kinetic energy
$\epsilon_{\vec k}$ and an exchange interaction $J_{\vec k}$, where 
\def\dg{^\dagger} 
\begin{equation}
H[\vec A] =
\sum_{\vec k}\bigl[
\psi\dg_{\vec k} 
\epsilon_{(\vec k - \vec A \hat \omega)}\psi_{\vec k} 
+\frac {1}{2}
\vec \sigma_{-\vec k}
\cdot 
J_{(\vec k   - \vec A \hat \omega)}
\vec \sigma_{\vec k}\bigr],
\end{equation}
where $\vec \sigma_{\vec k }$ is the Fourier transform of the spin density
at wavevector
$\vec k$ .
Differentiating with respect to $\vec A$ gives 
\begin{equation}
\vec j =i  \sum_{\vec k}
\biggl[
\bigl(\vec\nabla_{\vec k}\epsilon_{\vec k})
\psi\dg_{\vec k} 
\!\stackrel{\leftrightarrow}{\partial_t}\!
\psi_{\vec k}
+
\frac{1}{2}\bigl(\vec \nabla_{\vec k} J_{\vec k}\bigr) \vec
\sigma _{-\vec k}\cdot
\!\stackrel{\leftrightarrow}{\partial_t}\!
\vec \sigma_{\vec k}
\biggr].\end{equation}
The second term reflects the additional
heat flow created  by the exchange interaction. In an antiferromagnetic
Mott insulator, it is this term which is responsible for the
conduction of heat by spin-waves.  The approach illustrated
here can be used on any Hamiltonian,
without the need to develop the equations of motion for that
particular model.
Past efforts to compute the thermal current contribution from
interactions have adopted an equation of motion approach on a case-by-case
basis. 
This yields long expressions for the thermal current, where the time derivatives 
are expanded in their full glory.\cite{mahan,polaron}

The usefulness of the gauge-theoretic derivation of the thermal current
lies partly in its invariance
under the renormalization group. This means that it can 
be directly applied to the effective Lagrangian that describes
the low energy physics of an interacting system.  In any case where the
low energy physics is described by weakly interacting quasiparticles
with energy $E_{\vec k}$, application of the same procedure yields
\begin{equation}
\vec j = i\sum_{\vec k} \bigl(\nabla_{\vec k}E_{\vec k}\bigr)a\dg_{\vec k} 
\stackrel{\leftrightarrow}{\partial_t} 
a_{\vec k} = \sum_{\vec k}\bigl( E_{\vec k}\nabla_{\vec k}E_{\vec k}\bigr) n_{\vec k},
\end{equation}
where $n_{\vec k}= a\dg_{\vec k}a_{\vec k}$ is the quasiparticle number
operator and we have used the equation of motion to make the last substitution.
When the mass renormalization of the electrons is highly anisotropic,
the quasiparticle thermal current contains  a large interaction component.
Microscopic calculations that ignore these 
contributions fail to recover the correct quasiparticle description
of the heat current at low temperatures.

A particularly important illustration 
of this effect  occurs in an
anisotropic superconductor.  In this case,   the BCS Hamiltonian
takes the form
\begin{eqnarray}
H[\vec A ] &=& {1/2}
\sum_{\vec k}\Psi\dg_{\vec k}{\cal E}_{{\vec k} - \vec A \hat \omega }
\Psi_{\vec k}, \nonumber \\
{\cal E}_{\vec k} &=&  
\left( 
\begin{array}{cc}
\epsilon_{\vk}\hat{1} & \hat{\Delta}_{\vk} \\
\hat{\Delta}^{\dagger}_{\vk} & -\epsilon_{\vk}\hat{1},
\end{array} \right)
\end{eqnarray}
where  $\hat{\Delta}_{\vec k}$
is the anisotropic gap function,
$\Psi\dg_{\vec k}$ 
is the four-component Nambu spinor for the electrons.
The thermal current operator is then 
\begin{equation}
\vec j =
{ 1\over 2}\sum_{\vec k}\Psi\dg_{\vec k}
\nabla_{\vec k}{\cal E}_{\vec k}\hat \omega
\Psi_{\vec k}.
\end{equation}
The off-diagonal terms in this expression  derive
from the interaction contribution to the thermal current.
In the presence of weak scattering, we can use the equation of motion
to replace
\begin{equation}
\nabla_{\vec k}{\cal E}_{\vec k}\hat \omega \rightarrow
\frac{1}{2}\{
{\cal E}_{\vec k}, 
\nabla_{\vec k}{\cal E}_{\vec k}\} =  E_{\vec k} 
\nabla_{\vec k} E_{\vec k}\underline{1},
\end{equation}
where $E_{\vec k}
= (\epsilon_{\vec k}^2 + \Delta_{\vec k} ^2)^{1/2}$. This 
recovers the quasiparticle form (9). Note however, that the
Fermi velocity is replaced by the 
the quasiparticle group velocity
$\vec {\cal V}_{\vec k}=\nabla_{\vec k} E_{\vec k}$, which
has important components {\em parallel} to the Fermi surface. 
These can significantly 
affect the anisotropy of the thermal conductivity.  
\begin{figure}[btp]
\centerline{\psfig{file=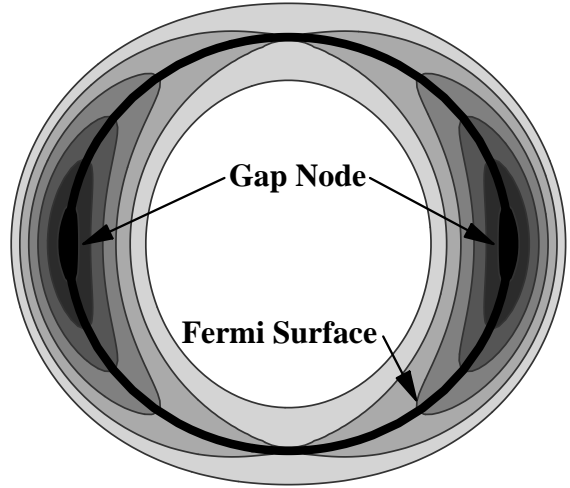,width=3.0in}}
\vskip -0.4truein
\caption{Constant energy contours surrounding a line 
node in a ``polar'' superconductor.  The thermal current operator is normal to the energy contours and points radially outwards from the gap node.
The dense line indicates position of 
Fermi surface and the dark region represents the
line node, which lies perpendicular to the plane of the paper.
}
\label{fig1}
\end{figure}
We now use (11) and apply the Kubo formula, to obtain
\begin{eqnarray}
\kappa^{ij}= \frac{\pi}{2 T} \sum_{\vk}
\int^{\infty}_{-\infty} {d\omega} \omega^2
\left( -\frac{\partial f}{\partial \omega}\right)  
{\Lambda^{ij}(\vec k, \omega)},
\label{thcond}
\end{eqnarray}
where \begin{equation}
\Lambda^{ij}(\vec k, \omega) =
Re \left\{ Tr\! \left[ 
({\nabla_{k}}\hat{{\cal E}_{\vec k}})^i
{\rm A}_{\vec k}(\omega)
( {\nabla_{k}}\hat{{\cal E}_{\vec k}})^j {\rm A}_{\vec k}(\omega)
\right]\right\}\ ,
\end{equation}
${\rm A}_{\vec k}(\omega)$ is the matrix spectral function
and $f$ is the  Fermi function.
Carrying out the trace we find
\begin{equation}
\kappa^{ij} = { 1 \over 2T} \int _{-\infty}^{\infty}
d \omega
\omega^2\biggl(
-{d f \over d\omega}
\biggr){ N(\omega) \over \Gamma(\omega)}
\langle \vec {\cal V}^i\vec {\cal V}^j \rangle_{\omega}
,
\end{equation}
where $\Gamma(\omega)$ is the quasiparticle scattering rate
and
\begin{equation}
N(\omega)\langle \vec {\cal V}^i\vec {\cal V}^j\rangle_{\omega}
 =\sum_{\vec k}
\vec {\cal V}_{\vec k}^i
\vec {\cal V}_{\vec k}^j \delta(\omega- \vec E_{\vec k})
\end{equation}
describes the quasiparticle velocity distribution.

To show how the interaction contribution to the thermal current
modifies the thermal conductivity, 
consider the case of a ``polar'' superconductor, with a spherical
Fermi surface and gap function 
$|\Delta_{\vk}|^{2}=\Delta_{0}^{2} \hat{k}_{z}^{2}$, which leads to
a line of gap nodes in the basal plane $k_z=0$.
The kinetic contribution to the thermal current operator always
lies normal to the electron Fermi surface, and if we omit the
effects of interaction, the low temperature thermal conductivity
lies predominantly in the basal plane.  In the  correct calculation,
the current carried by each excitation lies in the direction of the
quasiparticle velocity $\vec v^{QP}_{\vec k}$, which 
lies normal to the surfaces of constant energy (Fig. 1). 
The line-node in the basal plane is surrounded by elliptical surfaces
of constant energy and consequently, in the vicinity of the gap node
there are quasiparticles which propagate perpendicular to the basal
plane, giving rise to a increased thermal conductivity in this direction.
At low temperatures the 
thermal conductivity in both directions 
is proportional to $T^2$, with a finite ratio at $T=0$
\begin{equation}
{\kappa^{zz} \over \kappa^{xx}} = \frac {1}{2}\left( {\Delta_o 
\over \epsilon_F}\right)^2.
\end{equation}
Without interaction effects, this ratio vanishes at $T=0$.
\begin{figure}[btp]
\psfig{file=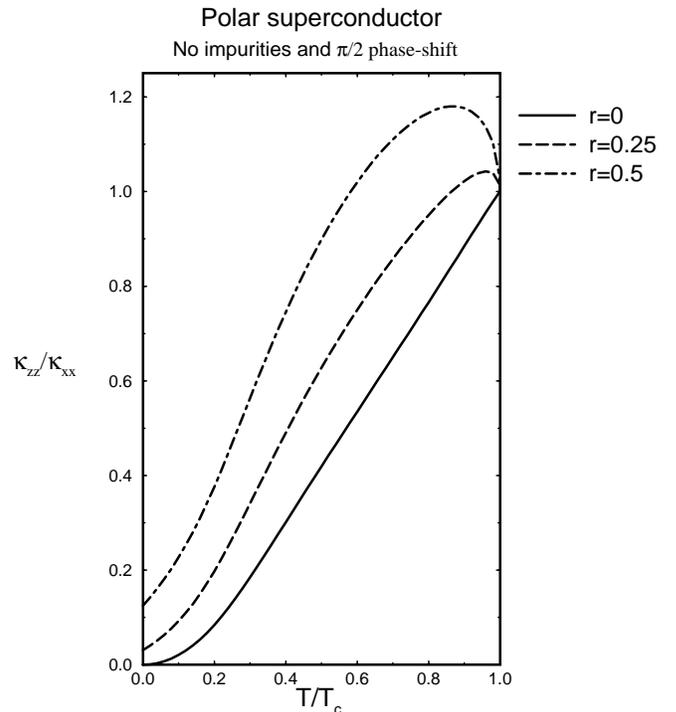,width=3.375in}
\caption{Ratio between the thermal conductivity perpendicular and
parallel to the basal plane as a function of the normalized
temperature for different values of the ratio
$r=\Delta_{0}/\epsilon_{F}$.}
\label{fig2}
\end{figure}

In Figure 2 we show how the anisotropy in the
thermal conductivity varies as the ratio between
the maximum gap and the Fermi energy
is increased from zero.  Two effects of interest are observed:
(i) the anisotropy is reduced by these interaction effects;
(ii) the thermal conductivity is enhanced in the region close to the
transition temperature. These effects become important
when the transition temperature
is a substantial fraction of the Fermi energy.  
Such considerations may be particularly relevant to heavy fermion
superconductors, of which two cases are particularly worthy of mention:
$UBe_{13}$ and $UPt_{3}$. In both cases,
NMR measurements suggest the presence of line nodes.
In the former case, 
superconductivity develops before a Fermi liquid regime
has been established, so it is very likely that $\Delta/\epsilon_F$ is
large.  In the latter case, the conventional wisdom  is that 
$\Delta/\epsilon_F\sim 0.05-0.1$. However, measurements of the thermal conductivity
on very pure samples of this material\cite{Taillefer} show that 
$\kappa^{zz}/\kappa^{xx}\sim {\ \rm constant}$ at low temperatures;
they also display a marked peak in the ratio $\kappa^{sc}/\kappa^n$ near
$T_c$.  Existing approaches \cite{recent} attribute these effects to a
combination of
both elastic and inelastic scattering.  While such an explanation is
surely feasible, an alternative explanation might be obtained
by assuming a larger 
ratio $\Delta/\epsilon_F$, and attributing these features to
the interaction contribution to the thermal current that has been
hitherto ignored.

The main point of this paper has been to emphasize that interactions
actually modify the thermal current, which can have important effects
on the thermal conductivity. As an illustration of these ideas,
we have considered the effect 
of gap anisotropy in polar superconductors. Our basic approach 
can be applied to
other gap anisotropic systems. One very interesting case
in this respect may be the small gap Kondo insulators,
where the insulating gap appears to exhibit a point node\cite{Miyake}.
Similar considerations can also be applied
to the momentum current, a point which is important in the interpretation
of measurements of the
ultrasound attenuation in strongly correlated systems. 
This could explain why early ultrasound measurements in $UBe_{13}$
\cite{Batlogg}
found no anisotropy, despite the clear suggestion of line nodes from
NMR measurements.\cite{maclaughlin} 

A secondary aspect of our work concerns the 
observation that thermal currents can be treated using a gauge theoretic
approach.  Just as momentum currents, or stress  ($\sigma$)
in a material result
from the gauge field we call strain ($u$), $\sigma = R u$, 
energy currents result from
a ``temporal strain'' which we represent by the gauge field $\vec A$.
If we calculate  the heat current response to this fictitious field
\begin{equation}
\vec j_T(\omega) = - Q(\omega) \vec A(\omega),
\end{equation}
we find that the ``London Kernel'' is directly
related to the thermal conductivity by the
relation $Q(\omega) = i \omega T\kappa (\omega)$, from which we deduce that
\begin{equation}
{\partial \vec A \over \partial t} = -{ \vec \nabla T \over T}.
\end{equation}
In other words, $\vec A$ is the gauge field conjugate to thermal
gradients.
By analogy with the case of momentum currents,
the quantity $\vec A$ may be regarded as a sheer
in time. Just as broken spatial translation symmetry enables
a material to support a persistent stress 
or ``momentum superflow'',  we are led to speculate that if it were 
possible to produce a state of condensed matter with spontaneously
broken time translation symmetry, such a system would 
exhibit ``heat superflow'', manifested by an infinite thermal
conductivity.  

We are grateful to Micha Berkooz, Denis Golosov, Andres Jerez,
Gabriel Kotliar and Andrew Schofield for useful
discussions. This research was supported by NSF grant DMR-93-12138.

\end{document}